\documentclass[aps,pre,twocolumn]{revtex4-1}
\usepackage{amsmath,bm,epsfig,,subfigure}
\usepackage{color}
\usepackage[normalem]{ulem}

\def\Fbox#1{\vskip1ex\hbox to 8.5cm{\hfil\fboxsep0.3cm\fbox{%
  \parbox{8.0cm}{#1}}\hfil}\vskip1ex\noindent}  

\newcommand{\B}[1]{{\bm{#1}}}
\let \= \equiv \let\*\cdot \let\~\widetilde \let\-\overline

\begin{document}
\title{On the Effect of Micro-alloying on the Mechanical Properties of Metallic Glasses}
\author{Oleg Gendelman$^1$, Ashwin J$^2$, Pankaj Mishra$^2$,  Itamar Procaccia$^2$ and Konrad Samwer$^3$}
\affiliation{$^1$ Faculty of Mechanical Engineering, Technion, Haifa 32000, Israel, $^2$ Dept. of Chemical Physics,
The Weizmann Institute of Science, Rehovot 76100, Israel, \\$^3$ I.Physikalisches Institut,
Universitaet G\"ottingen, Friedrich-Hund-Platz 137077  G\"ottingen, Germany.}
\begin{abstract}
``Micro-alloying", referring to the addition of small concentration of a foreign metal to a given metallic glass,
was used extensively in recent years to attempt to improve the mechanical properties of the latter. The results
are haphazard and nonsystematic. In this paper we provide a microscopic theory of the effect of micro-alloying,
exposing the delicate consequences of this procedure and the large parameter space which needs to be controlled.
In particular we consider two very similar models which exhibit opposite trends for the change of the shear
modulus, and explain the origins of the difference as displayed in the different microscopic structure and properties.
\end{abstract}
\maketitle

\section{Introduction}

Bulk Metallic Glasses have attracted considerable attention due to their high strength compared to the crystalline
counterparts \cite{07Wan}. On the other hands these promising materials exhibit a catastrophic brittle fracture and strain softening; these undesired properties seriously limit their uses as engineering materials \cite{09CDES,11DFCSGJ}. In recent years many laboratories tried to improve the mechanical properties of metallic glasses by adding small concentrations of a foreign metal \cite{07HDJS,12GDCJ}. Theoretically an attempt was made to explain the effect of micro-alloying by
adding pinned particles to the glass forming system \cite{13DMPS}; such a procedure can only increase the observed
shear modulus as well as the toughness.
It turns out that in experiments the actual effect of ``micro-alloying" is hardly predictable, and large efforts are necessary to try out different additives at different conditions with haphazard results concerning the observed mechanical
properties. Thus for example in Ref. \cite{13ZWZSCWLL} one found a decrease in the mechanical modulus whereas
in Ref. \cite{11HCMX} the opposite was found. The aim of this paper is to go beyond the ideal model of pinning, and to understand, based on a microscopic theory, the origin of these highly non-universal consequences of micro-alloying.

For simplicity and concreteness we will focus in this paper on simple model glasses at zero temperature. The reason for this
choice is that at $T=0$ and for quasistatic strain protocols we possess an exact theory for the mechanical properties
and in particular the shear modulus that we analyze below \cite{10KLP,11HKLP}. The exact theory allows us to probe the precise reasons for
the changes in shear modulus upon the addition of the foreign particles, exposing the very large parameter space
that needs to be controlled. Choosing randomly foreign particles whose interactions with the present ones in a given
glass are not precisely characterized can lead to changes in the shear modulus that are highly unpredictable.
In Sect. \ref{models} we present the two models used in this paper. Sect. \ref{results} presents the results of micro-alloying
in terms of the observed stress vs. strain curves and the shear modulus. Sect. \ref{theory} discusses the microscopic
theory and explains the observed results. In Sect. \ref{conc} we offer a summary and conclusions.

\section{The Models}
\label{models}

\subsection{The basic model glass}

As our model glass (before micro-alloying) we select the well-studied \cite{13DMPS} model of a
binary $50-50$  Lenard-Jones mixture whose potential energy for a pair of particles labeled  $i$ and $j$ has the form

\begin{eqnarray}
&&U_{ij}(r_{ij}) = 4\epsilon_{ij}\Big[\Big(\frac{\sigma_{ij}}{r_{ij}}\Big)^{12} - \Big(\frac{\sigma_{ij}}{r_{ij}}\Big)^{6} + A_0\nonumber\\&&+ A_2\Big(\frac{r_{ij}}{\sigma_{ij}}\Big)^2+A_4\Big(\frac{r_{ij}}{\sigma_{ij}}\Big)^4+A_6\Big(\frac{r_{ij}}{\sigma_{ij}}\Big)^6\Big] \ .
\end{eqnarray}
Depending on whether particles $i$ and $j$ are `small' (S) or `large' (L), the length parameters $\sigma_{ij}$ take on the values $\sigma_{_{SS}}$, $\sigma_{_{LL}}$ and $\sigma_{_{SL}}$, chosen to be $2\sin(\pi/10)$, $2\sin(\pi/5)$ and $1$ respectively. $\sigma_{_{SL}}$ is also referred to as simply $\sigma$ and it acts as the fundamental scale. Thus the potential is cut-off at $r/\sigma = 2.5$. We guarantee that this cut-off is smooth with two smooth derivatives, and this is
the function of the parameters $A_0$, $A_i$ (i =2, 4, 6). Finally the energy parameters $\epsilon_{_{SS}} = \epsilon_{_{LL}} = 0.5, \epsilon_{_{SL}} = \epsilon_{_{LS}}=1$. Below $\epsilon_{_{SL}}$ acts as the energy scale in units for which the
Boltzmann constant is unity.

\subsection{Modeling micro-alloying}

In order to model micro-alloying we replace a small percentage of `small' particles by marked particles, designated below as `M'. Then the mixture contains Small, Large and Marked particles. Importantly this substitution is made already in the
liquid state before the quench to an amorphous solid. We believe that this is in accordance with the laboratory procedure
of micro-alloying. Note that we are not trying to model a particular experiment of micro-alloying, but rather understand
the observed high sensitivity to particular choices of added metals.  Obviously, one can choose the marked particles in many ways, each contributing to a change in the mechanical properties of the mixture. For example, we can choose any of the interaction length-scales $\sigma_{_{MM}},~\sigma_{_{ML}}, \sigma_{_{MS}}$ differently, as well as the corresponding energy parameters. This opens up a six parameter phase space even for the very simple models that we discuss here. Other possibilities include 3-particle interactions, angle-dependent interactions, and what not. For concreteness we will examine only two models, not touching the range of interaction, taking for the ranges the same values for `M' as the small particle `S' that it replaces. The
only difference will be in the energy parameters that determine the depth of the potential. In both models considered
below we preform a standard quenching protocol from the melt to the amorphous solids at $T=0$. The quench of a liquid
with 5625 particles started at
temperature $T=1.2$ at a rate of $3.2\times 10^{-6}$. All the simulations are done in NVT ensemble with a density $\rho=0.976$.

{\bf Model A:} The presence of marked particles increases the potential depth between the marked and the small particles. Now the interaction parameter between the marked and glass forming particles have the values: $\epsilon_{_{MM}} = 0.5$, $\epsilon_{_{SM}} = 2.5$, $\epsilon_{_{ML}} = \epsilon_{_{LS}}=1$. The other parameters remain same as original glass forming mixtures. With this choice of energy parameters there will be a tendency of small particles to aggregate and cluster
around the M particles.

{\bf Model B:} In the second model it is assumed that marked particle attracts the large particles present in the mixture. To take it into account we have chosen the interaction parameters as $\epsilon_{_{MM}} = 2.5$, $\epsilon_{_{LM}} = 5.0$, $\epsilon_{_{MS}} = \epsilon_{_{LS}}=1$. Note that here the choice means that large particles will aggregate and cluster
around the M particles.

While these difference in choices seems slight, we will see below that they lead to strong and opposite trends in changing
the mechanical properties of the resulting glass.
\begin{figure}[h]
\includegraphics[scale = 0.32]{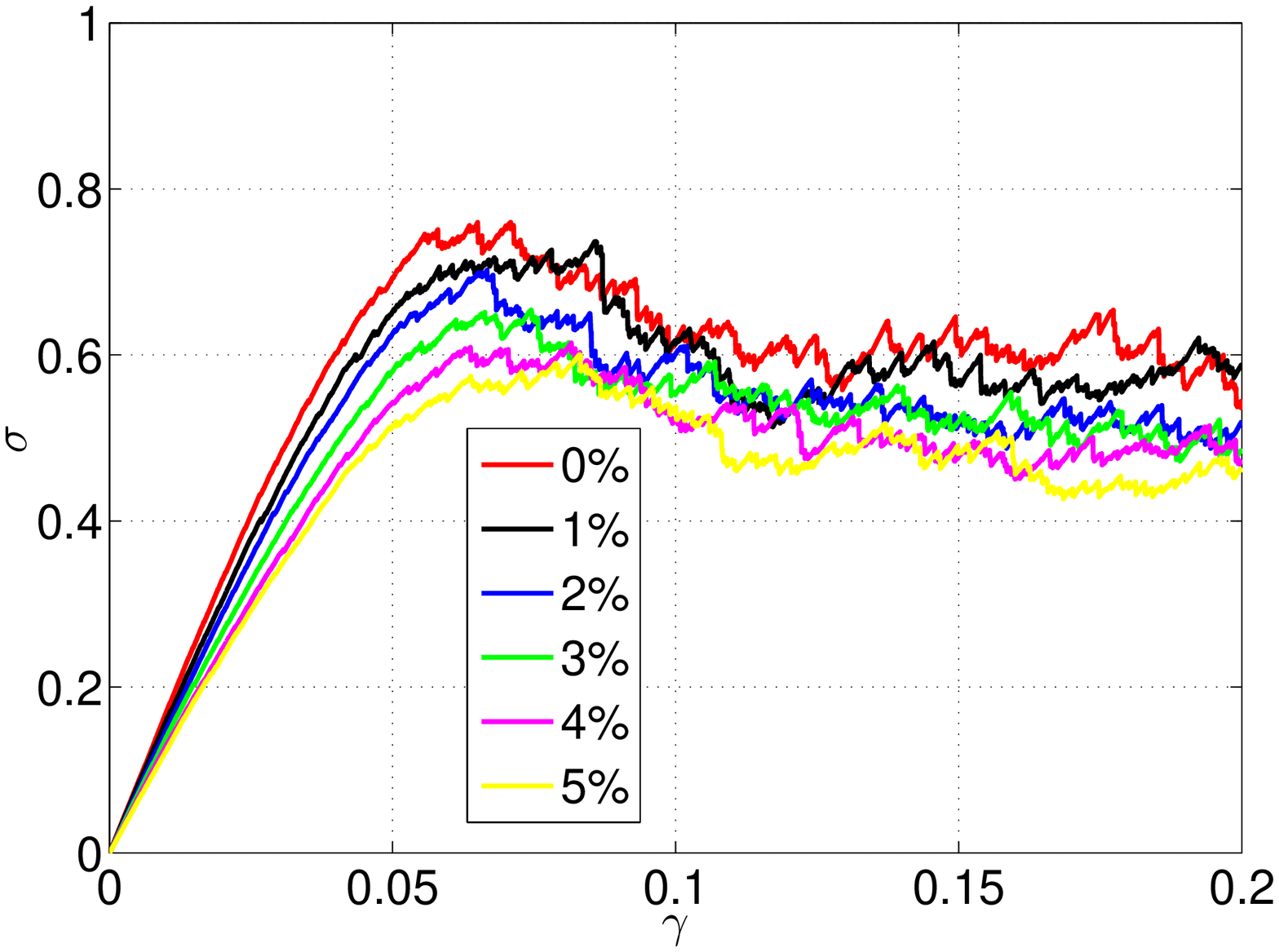}
\includegraphics[scale = 0.35]{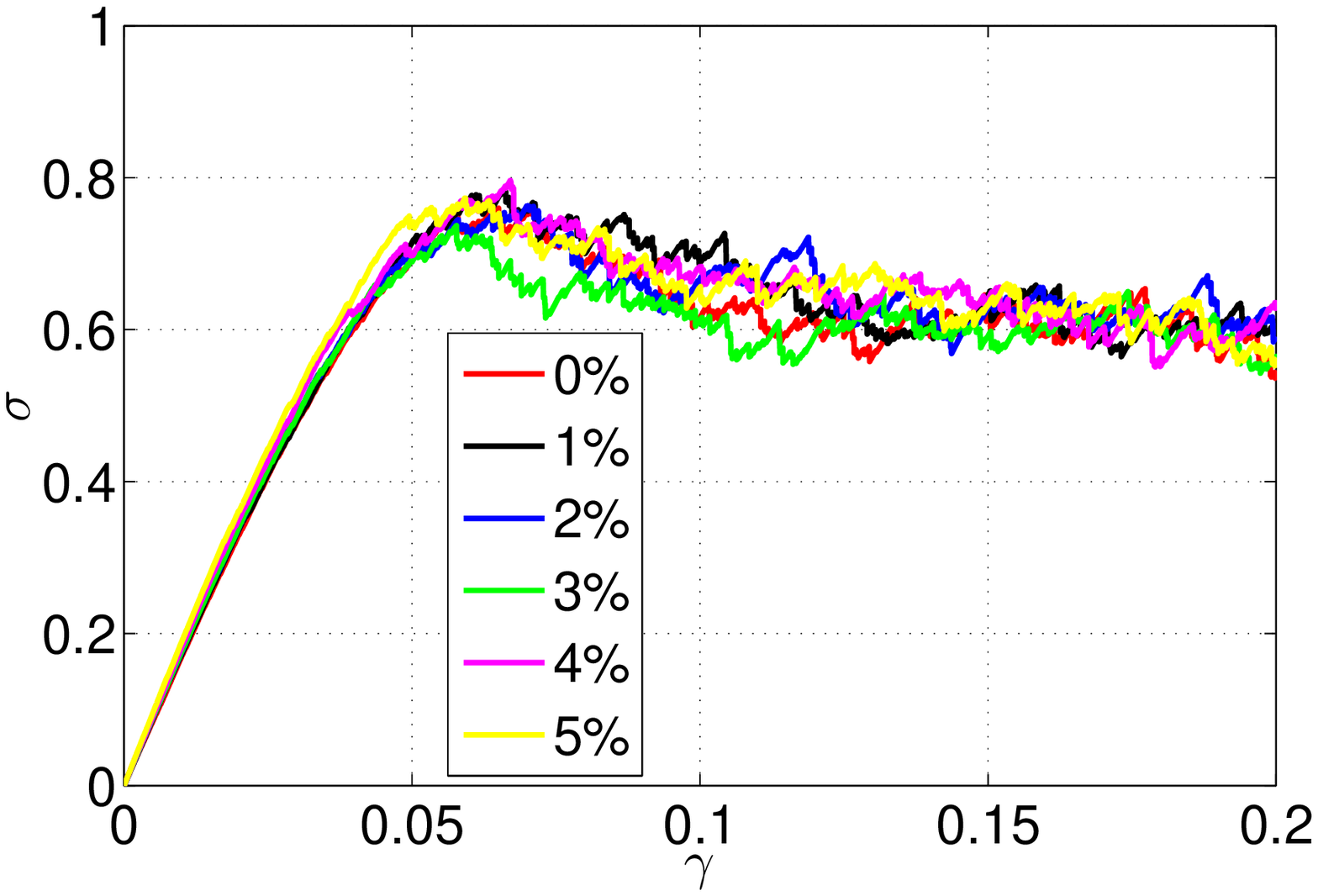}
\caption{Stress vs. strain curves in a quasi-static strain simulation for model A (upper panel)
and model B (lower panel). In model A we find a strong decrease in shear modulus, in model B a less pronounced
increase.}
\label{stvsst}
\end{figure}

\section{Observed Results of Micro-alloying}
\label{results}

The mechanical properties of the resulting mixtures are observed by straining the material at $T=0$ under quasi-static
shear. This means that after every infinitesimal change in the external strain $\gamma$ the system is relaxed by
gradient energy minimization to the nearest inherent state. The raw available data are the stress vs. strain curve
as shown in Fig. \ref{stvsst} for both models A and B at six  different concentrations of the marked particles M including
zero concentration.
\begin{figure}[h]
\includegraphics[scale = 0.25]{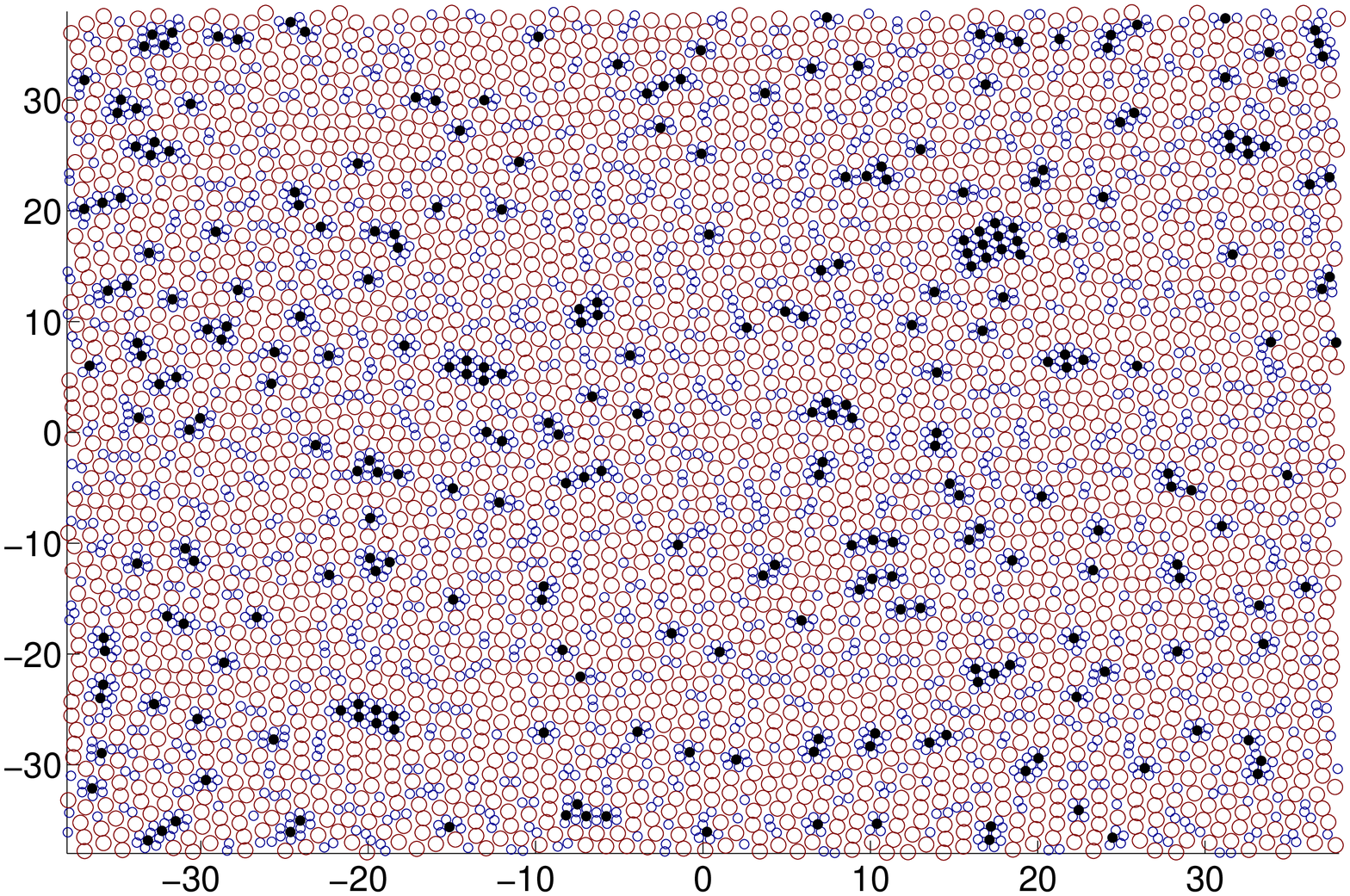}
\includegraphics[scale = 0.25]{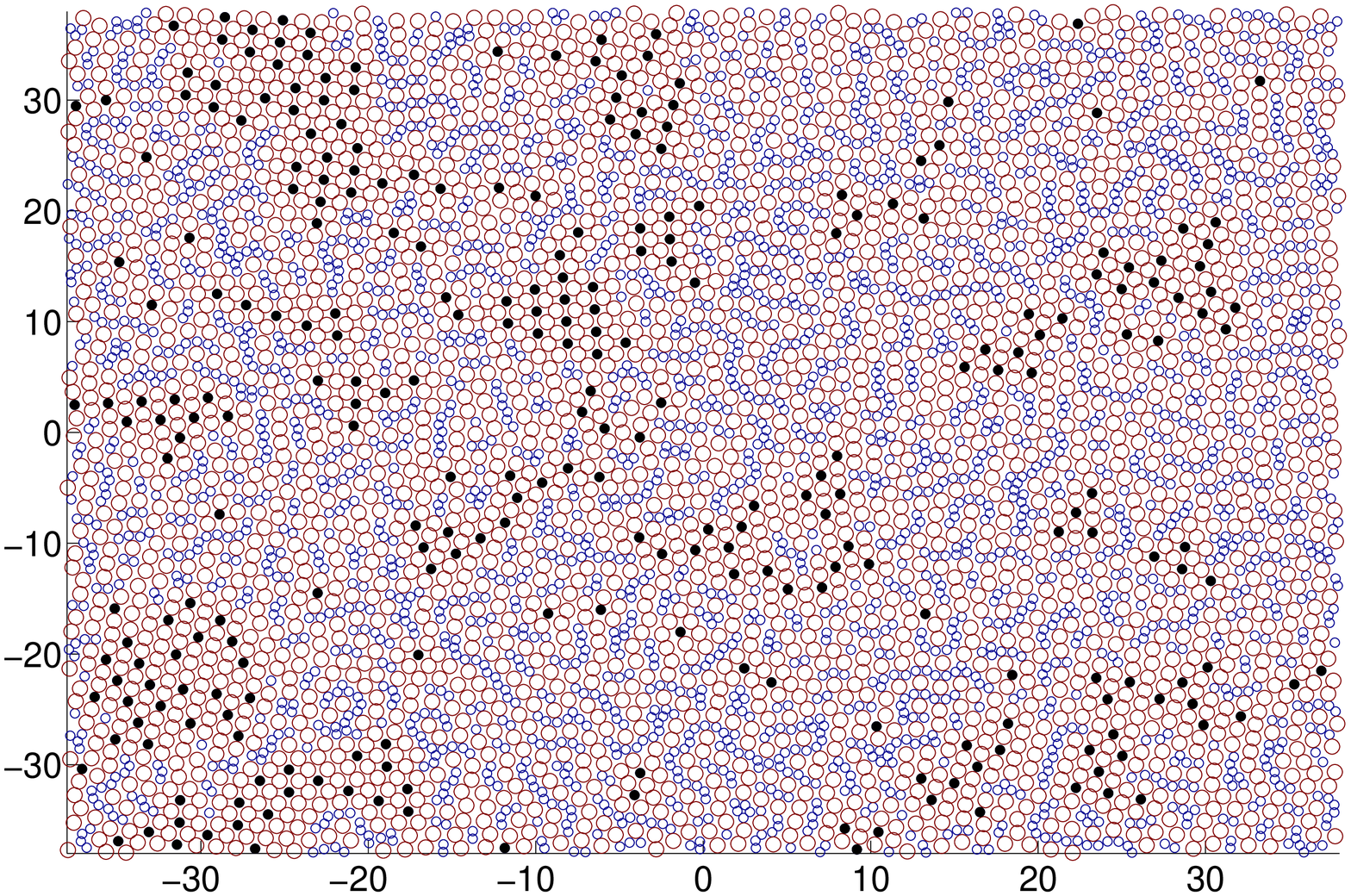}
\caption{The microscopic structure of the glasses at $T=0$ without external strain for 5\% of marked particles. In model A (upper panel) we see the aggregation of small particles (blue circles) around the marked ones (black dots), while in model B (lower panel) we observe large particles (red circles) aggregating around the marked particles.}
\label{struct}
\end{figure}
One observes readily that the shear modulus decreases significantly in model A, about 30\% with the addition of 5\%
M particles. In model B we observe the opposite trend, with about 12\% increase in shear modulus for the same
percentage of M particles.

\begin{figure}[h]
\includegraphics[scale = 0.25]{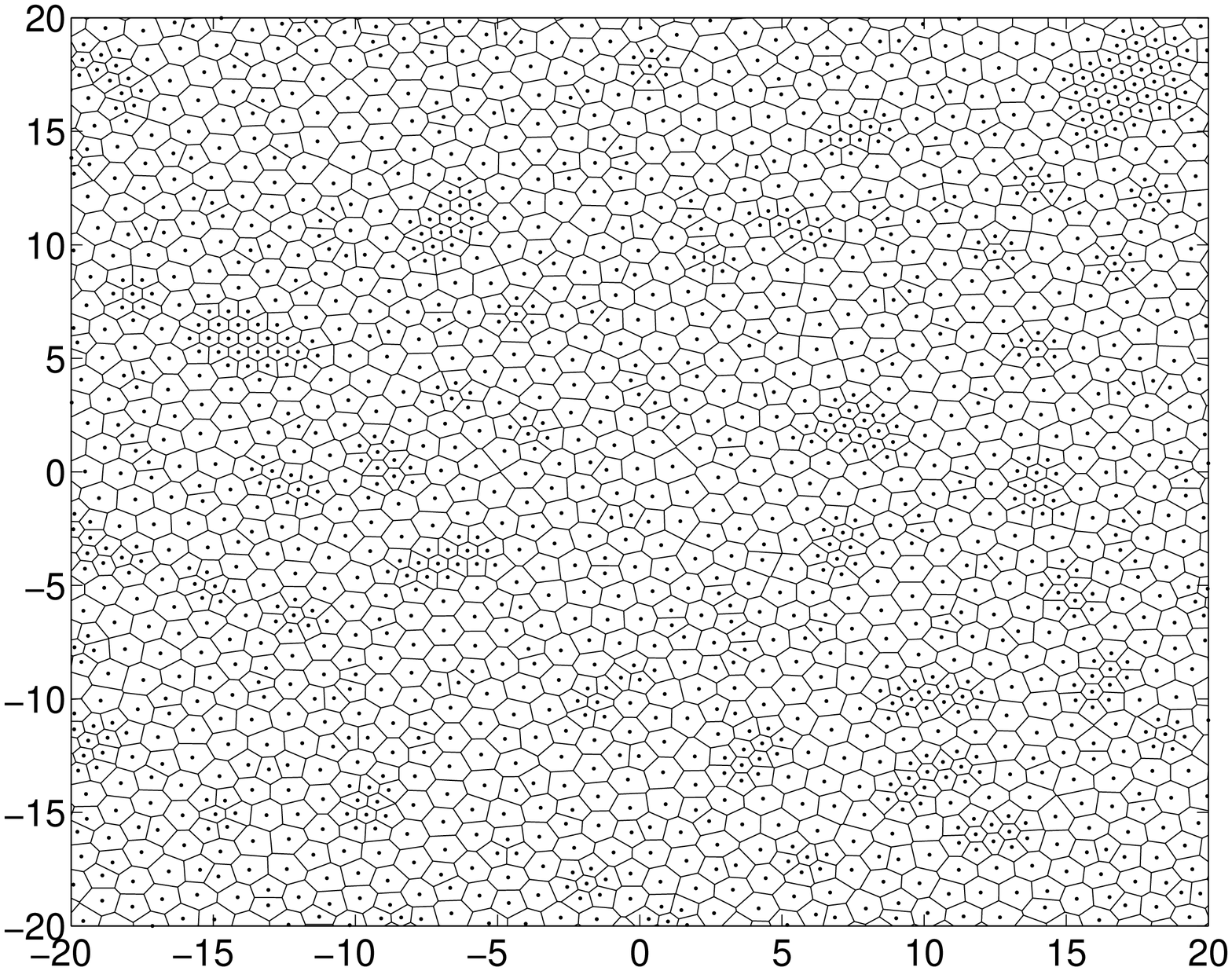}
\includegraphics[scale = 0.25]{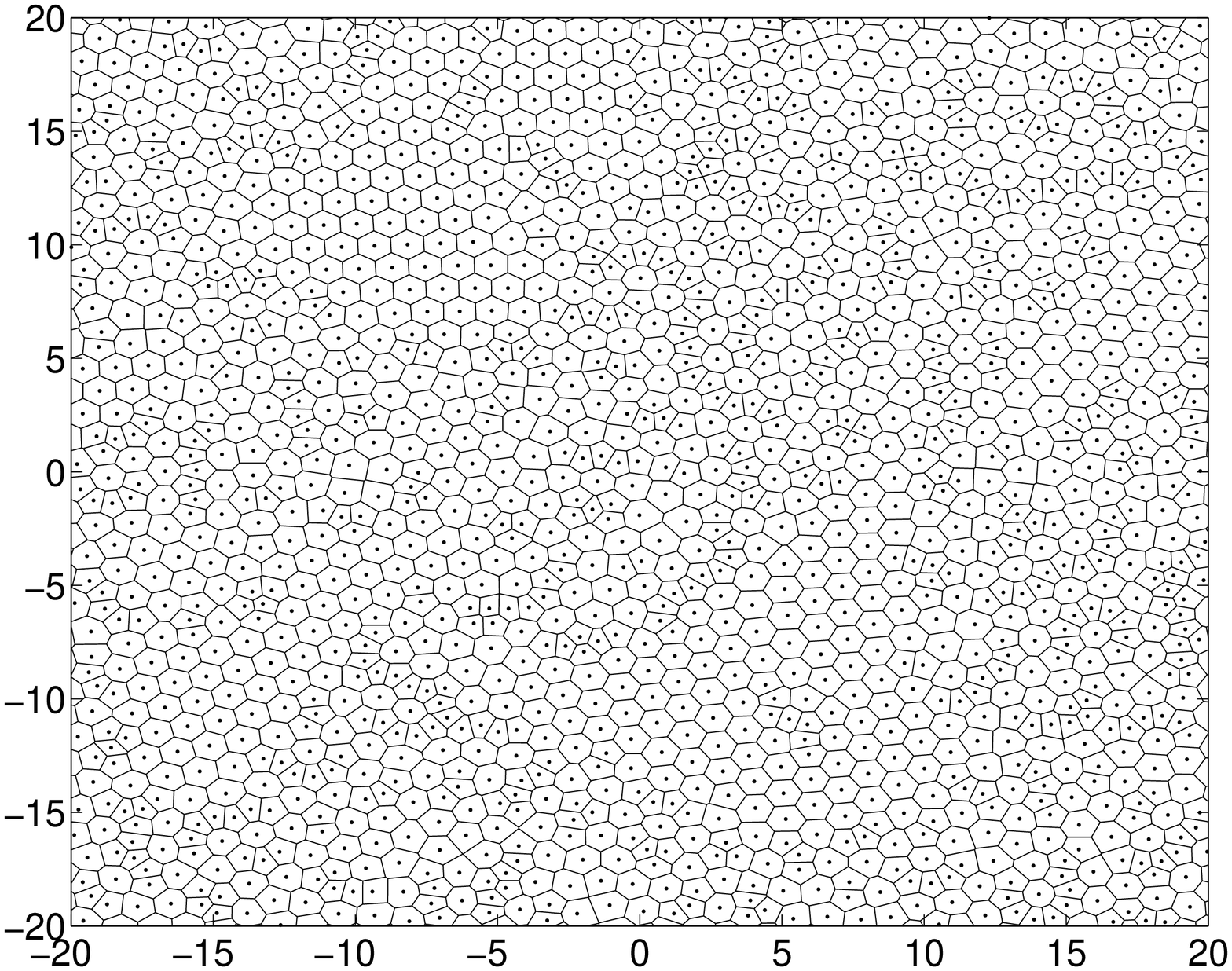}
\caption{The structure of the glasses at $T=0$ without external strain for 5\% of marked particles. Shown here is a
Voronoi map to stress the appearance of crystalline clusters for model A (upper panel) and model B (lower panel).}
\label{voronoi}
\end{figure}

One can attempt to figure out why these changes are occurring by looking directly at the structure of
the resulting glasses. These are shown for the two models in Fig.~\ref{struct} for a concentration of 5\% of marked particles. The images show clearly the tendency of clustering in the two models - in model A small particles aggregate around
the marked particles and in model B we have the opposite - large particles aggregate around the marked particles.
In both cases we see some crystalline order in these clusters. To make this fact more obvious we show in Fig. \ref{voronoi}
the same pictures but with a Voronoi map. This kind of clustering has been observed in experimental micro-alloying \cite{13ZWZSCWLL}, leading to a decrease in the Young modulus.
Realizing that in both cases we have clustering, then why the opposite tendencies
in the mechanical properties? To have an answer to this crucial question we need to turn to the microscopic
theory.

\section{Microscopic Theory}
\label{theory}

\subsection{General theory and results}
To understand the microscopic theory we need to recall what is entailed in
an athermal quasi-static (AQS) protocol to examine
the stress vs. strain curves \cite{13DMPS,10KLP}. In this procedure the particle positions in the system are first
changed by the affine transformation
\begin{equation}
x_i\to x_i+\delta \gamma y_i; \quad y_i\to y_i \ .
\end{equation}
This transformation results in the system not being in mechanical equilibrium, and we therefore allow
the non affine transformation $\B r_i\to \B r_i+\B u_i$. In this step the particles are moving, directed by a gradient energy minimization, to seek new positions that
annul the forces between all the particles. The existence of both affine and non-affine displacement results
in having two very different terms that determine the shear modulus. To see this recall that
the shear modulus is the second derivative of the energy of the system
with respect to the strain $\gamma$ \cite{10KLP}, i.e.
\begin{equation}
\mu = \frac{1}{V}\frac{d^2 U(\B r_1,\cdots,\B r_N;\gamma)}{d \gamma^2} \ .
\end{equation}
In our process the full derivative with respect to $\gamma$ translates to two contribution, one the direct
partial derivative with respect to $\gamma$ and the other, via the chain rule, the contribution due to the non-affine part of the transformation:
\begin{equation}
\frac{d}{d\gamma} = \frac{\partial}{\partial \gamma} + \sum_{i}\frac{\partial}{\partial \B u_i} \cdot \frac{\partial \B u_i}{\partial \gamma} \equiv  \frac{\partial}{\partial \gamma} +  \sum_{i}\frac{\partial}{\partial \B r_i} \cdot \frac{\partial \B u_i}{\partial \gamma}\ ,
\end{equation}
where the second equality follows from the form of the non-affine transformation where $d\B r_i=d\B u_i$.
To understand how to compute the constrained sum we need to recall that the force $\B f_i\equiv -\partial U/\partial \B r_i$ is zero before and after the affine and non-affine steps. Thus
\begin{equation}
-\frac{d}{d\gamma} \frac{\partial U}{\partial \B r_i} = 0= -\frac{\partial^2 U}{\partial \gamma\partial \B r_i} - \sum_{j}
\frac{\partial^2 U}{\partial \B r_i\partial \B r_j} \cdot \frac{\partial \B u_j}{\partial \gamma} \ .
\label{straightu}
\end{equation}
Denote now as usual the Hessian matrix $\B H$ and the non affine ``force" $\B \Xi$ \cite{10KLP}:
\begin{equation}
H_{ij} \equiv \frac{\partial^2 U(\B r_1,\cdots,\B r_N;\gamma)}{\partial \B r_i \partial \B r_j}\ , \quad
\Xi_i \equiv \frac{\partial^2 U(\B r_1,\cdots,\B r_N)}{\partial \gamma\partial \B r_i } \ .
\end{equation}
Inverting Eq. (\ref{straightu}) we find
\begin{equation}
\frac{d\B u_i}{d\gamma} = -\sum_{j} H^{-1}_{ij}\cdot \Xi_j \ .
\end{equation}

Applying these results to the definition of $\mu$ we end up with the exact expression
\begin{equation}
\mu=\frac{1}{V}\frac{\partial^2 U(\B r_1,\cdots,\B r_N;\gamma)}{\partial \gamma^2}-\frac{1}{V}
\sum_{i,j}\Xi_i\cdot  H_{ij}^{-1} \cdot \Xi_j \ , \label{defmu}
\end{equation}
where the first term is the well known Born contribution which we denote below as $\mu_B$. The second term exists only due to the non-affine
displacement $\B u_i$ and it includes the inverse Hessian matrix.
Needless to say, before we compute the non-affine contribution in Eq.~(\ref{defmu}) we need to remove the two Goldstone modes
with $\lambda=0$ which are the result of translation symmetry.

\begin{figure}[h]
\includegraphics[scale = 0.25]{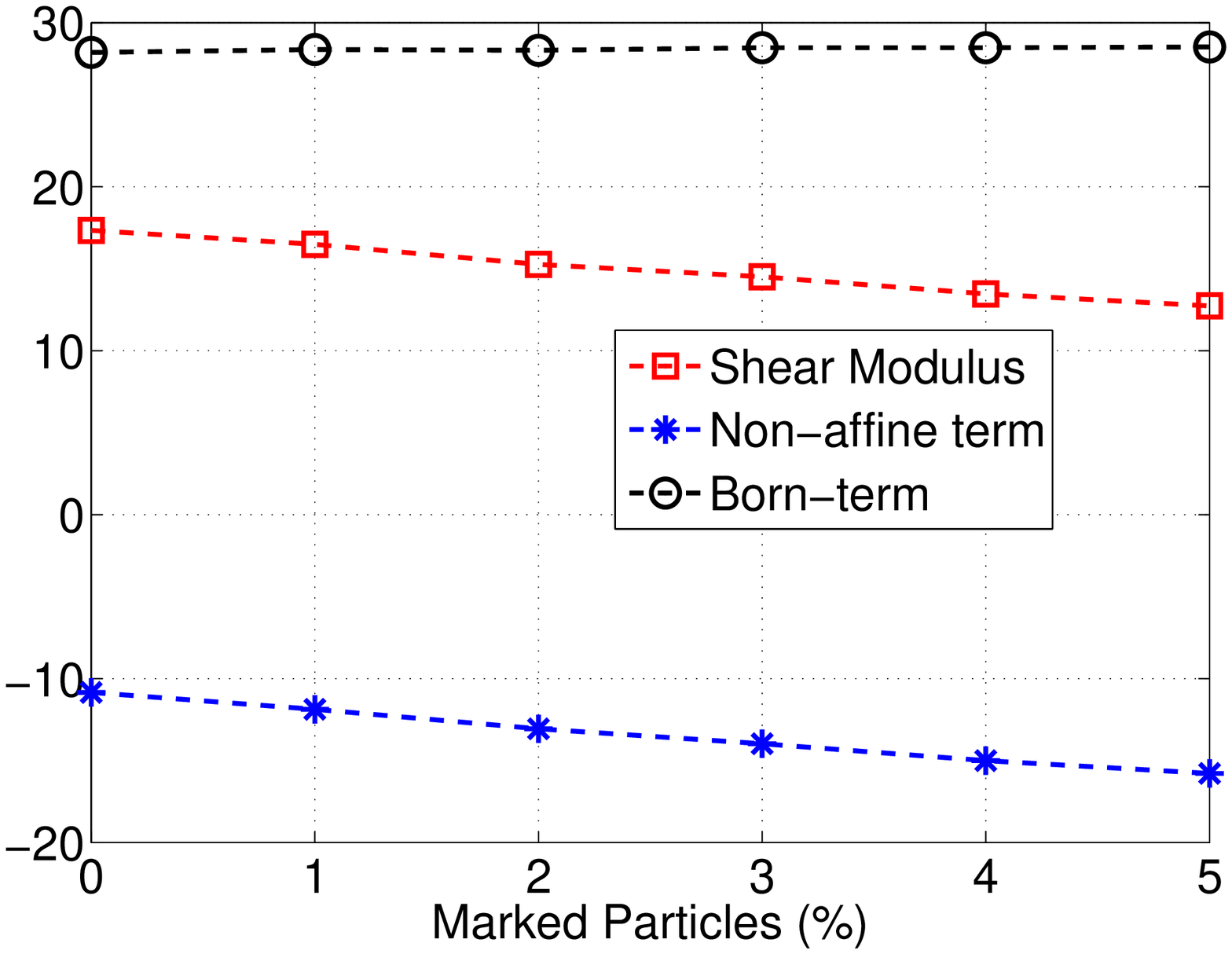}
\includegraphics[scale = 0.25]{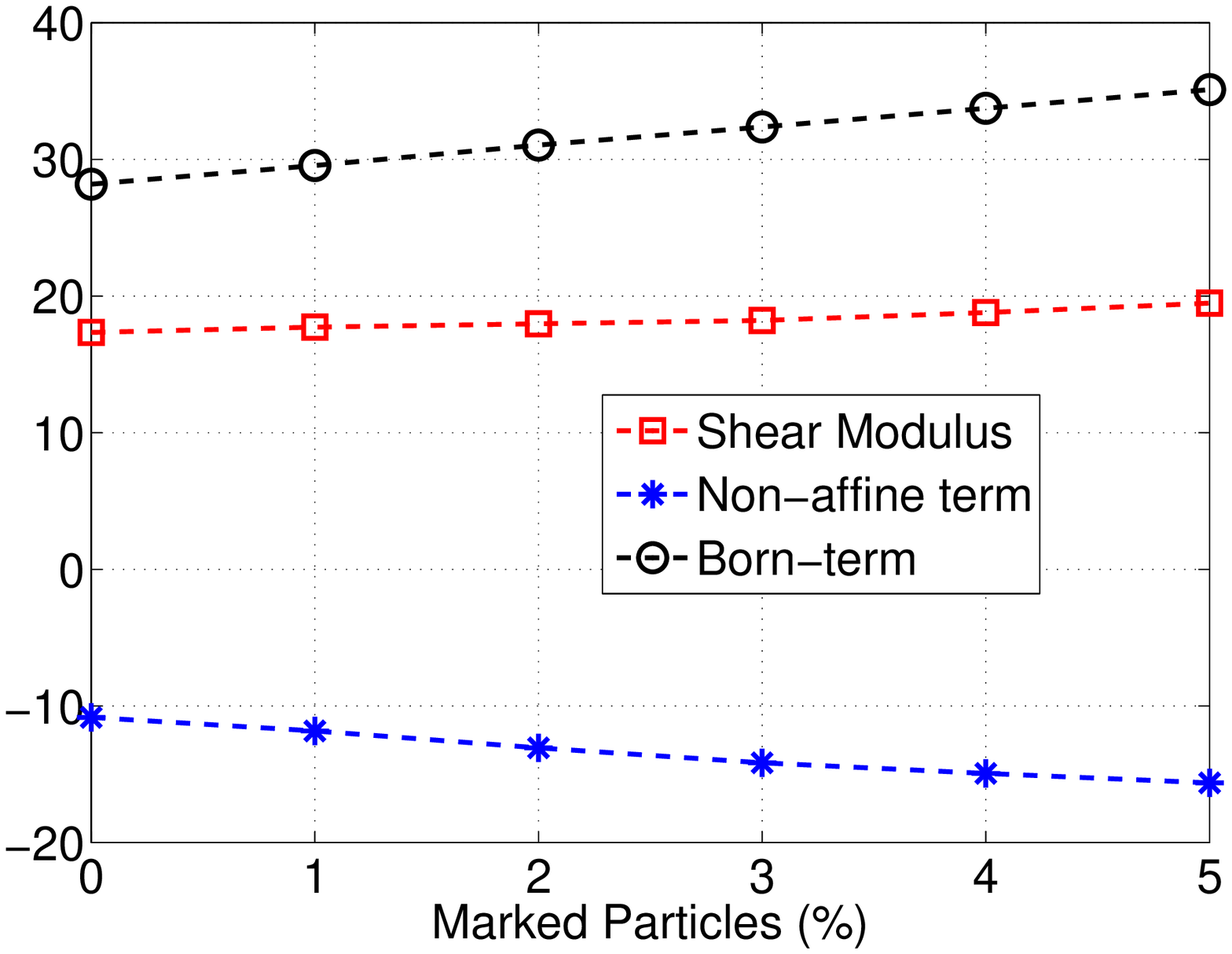}
\caption{The shear modulus of model A (upper panel) and model B (lower panel) in red line connecting the
square symbols. Also shown is the Born term in a black line connecting the round symbols and the non-affine
term in blue line connecting the star symbols. }
\label{mu}
\end{figure}

Both terms in Eq. (\ref{defmu}) are exactly computable, and indeed in Fig.~\ref{mu} we show the shear modulus in
models A and B, together with the two contributions that add up to the shear modulus, as a function of the concentration
of the marked particles.
The shear modulus itself is shown in red, connecting the square symbols. We see that it decreases as a function of the
concentration of the marked particles in model A and increases in model B. Also we exhibit in the same figure
the Born and the non-affine terms that are summed up to provide the shear modulus. In model A we see that the Born
term is hardly changing as a function of the concentration of M particles. Due to the significant increase in the non-affine
term, the overall shear modulus is a decreasing function of the concentration of the M particles. In contradistinction
in model B the Born term is sharply increasing as a function of the concentration of the M particles, and although
the non-affine part is again increasing, this is not enough to turn over the shear modulus, and it still keeps increasing
as a function of the concentration of the M particles. So what is the difference between the two models that is
responsible for these tendencies?

\subsection{Eigenvalues, eigenfunctions, density of states and the difference between the models}

The reasons for the difference between the two models become clearer when we examine in detail the properties of the
the Hessian matrix, its eigenvalues $\{\lambda\}$ and its eigenfunctions $\{\B\Psi_\lambda\}$ in the two models. First we consider the behavior of
 the Born term in the shear modulus, which is about constant in model A but increasing in model B. We relate the value of $\mu_B$ to the Debye cutoff which is the value of $\lambda$ where the eigenfunctions of the Hessian matrix get localized
 (Anderson localization). The relation is \cite{13ABP}:
\begin{equation}
\lambda_D = 8\pi \mu_B \ .
\end{equation}
The most direct way to find the Debye cutoff is by computing the participation ration of the eigenfunctions $\B \Psi_\lambda$.
This number is defined as
\begin{equation}
P(\B \Psi_\lambda) = \left[ \sum_{j=1}^N | \B \Psi^{(j)}_\lambda |^2 \right]^{-1}
\end{equation}
\begin{figure}[h]
\includegraphics[scale = 0.35]{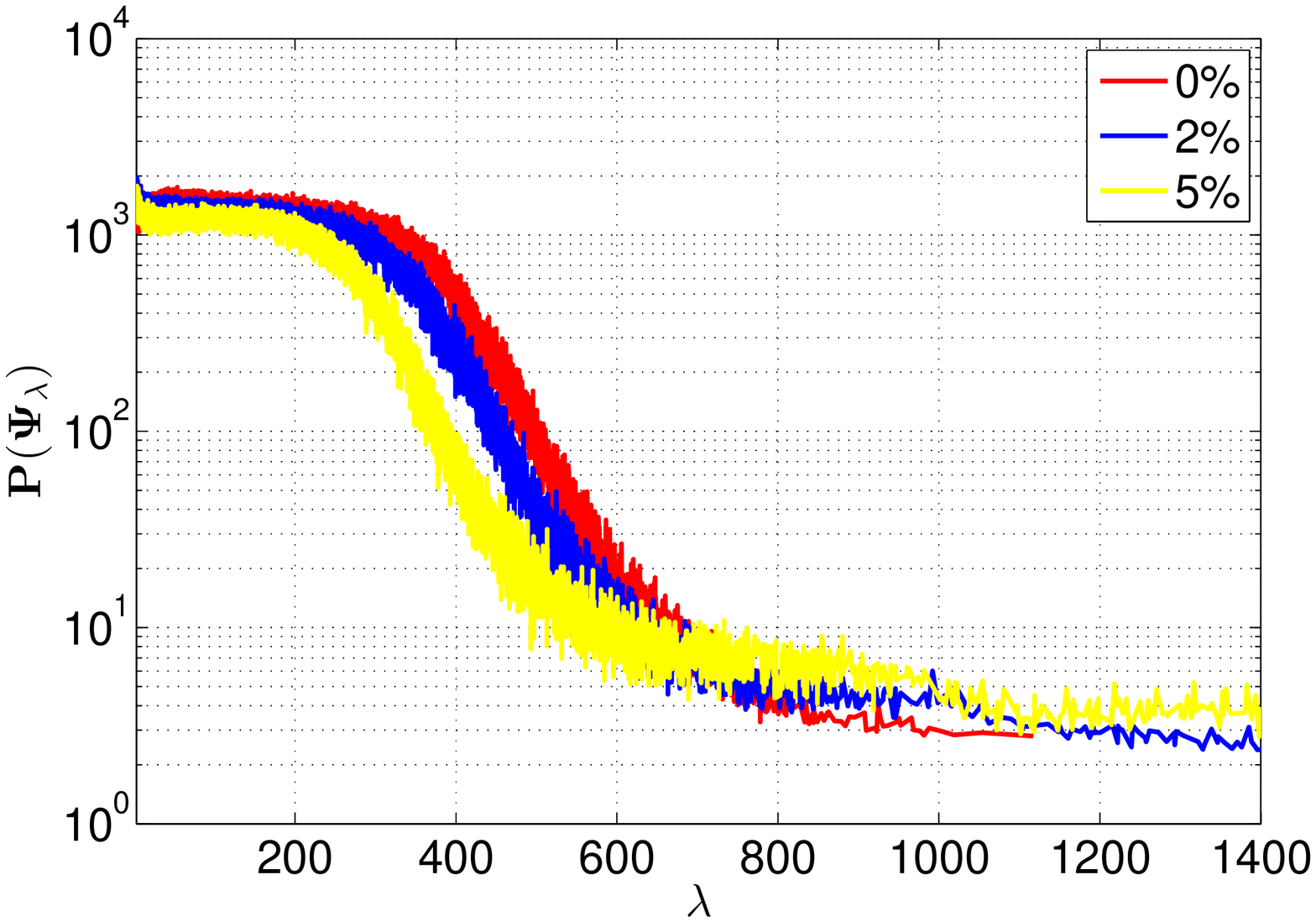}
\includegraphics[scale = 0.35]{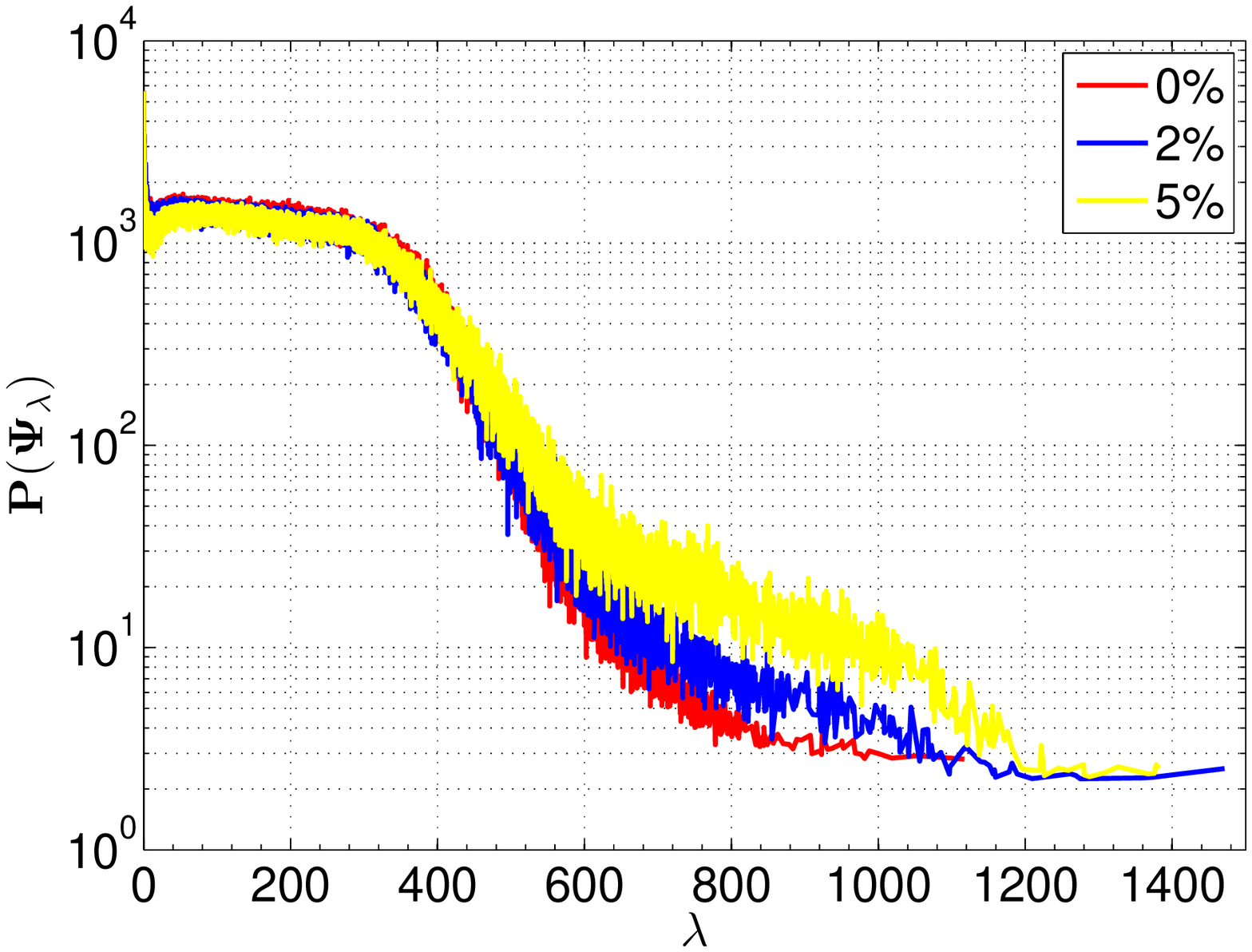}
\caption{The participation number of all the eigenfunctions of the Hessian matrix for model A (upper panel) and
model B(lower panel).}
\label{part}
\end{figure}
The participation numbers of all the eigenfunctions of the Hessian matrix are shown as a function of $\lambda$ in Fig.
\ref{part} for both models.
Indeed, the difference in behavior of the Born term $\mu_B$ between the two models becomes clear. While for
 model A the Debye cutoff remains the same, around $\lambda=600$, independent of the concentration of the marked particles,
it is not so for model B. Here there is a very significant increase in the Debye cutoff as a function of the concentration of the marked particles, from about $\lambda=600$ to about $\lambda=1100$. This is a direct explanation for the increase in the Born term in model B and the relative lack of increase in this term in model A. Of course this is also correlated with the increased crystalline clustering
in model B, an effect that must delay the Anderson localization of the eigenfunctions.

\begin{figure}[h]
\includegraphics[scale = 0.25]{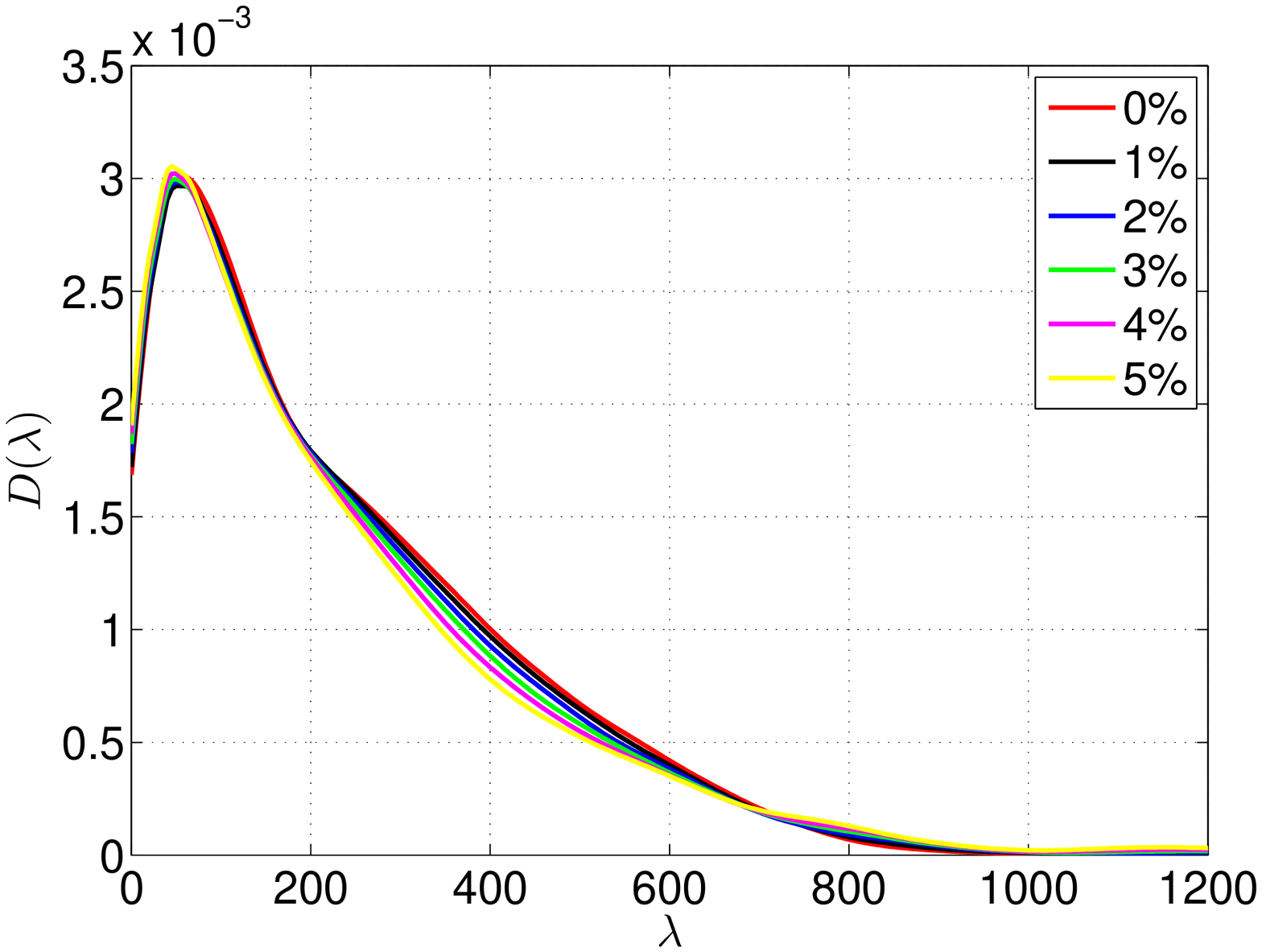}
\includegraphics[scale = 0.25]{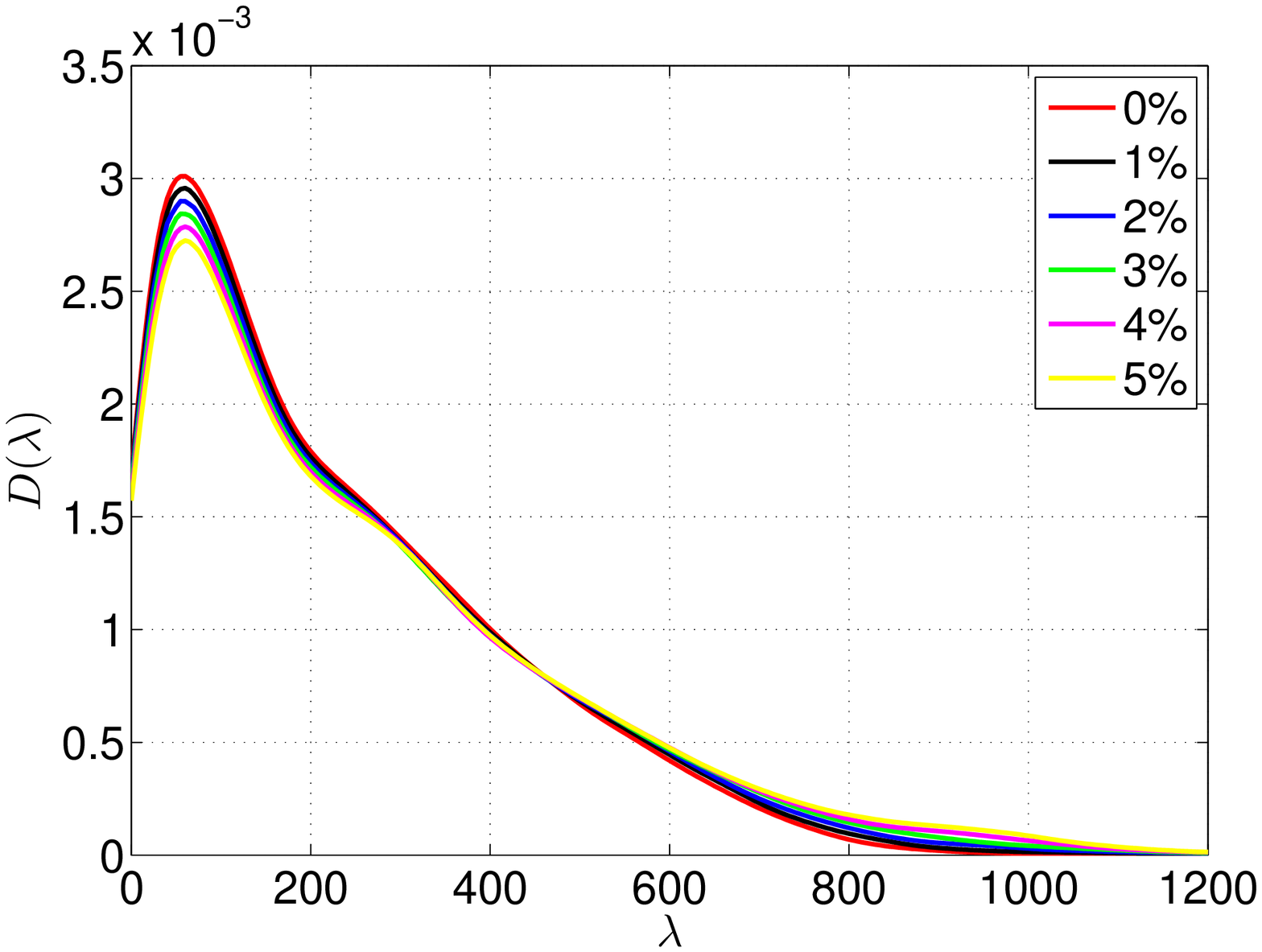}
\caption{The density of eigenvalues of the Hessian matrix for models A and B at different concentrations of the M
particles.}
\label{DOS}
\end{figure}

\begin{figure}[h]
\includegraphics[scale = 0.28]{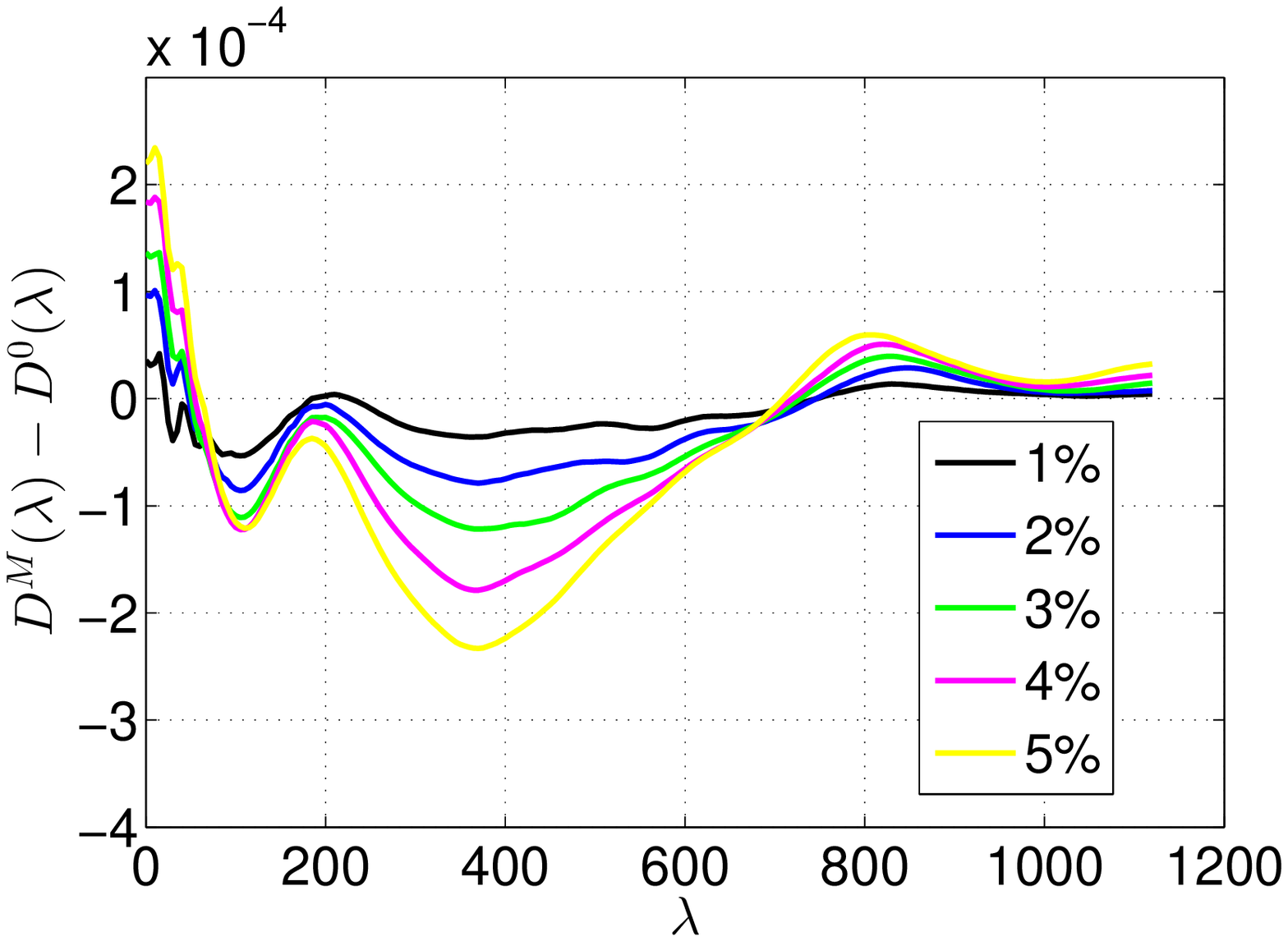}
\includegraphics[scale = 0.28]{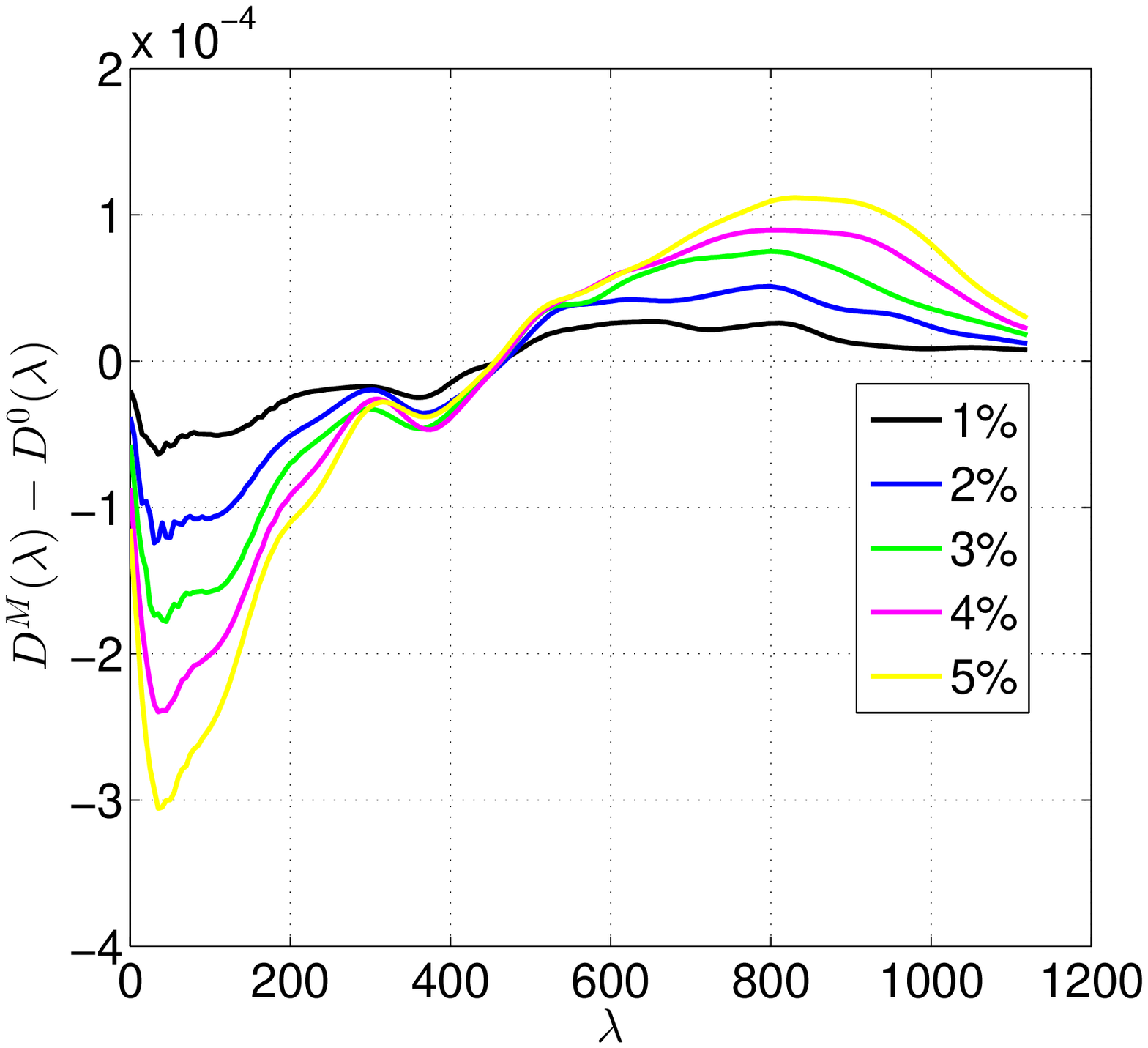}
\caption{The differences between the density of eigenvalues $D^M(\lambda)$ and $D^0(\lambda)$ for models A and B at different concentrations of the M particles.}
\label{diff}
\end{figure}

To shed additional light on both the Born and the non-affine term it is useful to consider the density
of states associated with the Hessian matrix. Denoting the eigenvalues of this matrix by $\{\lambda_i\}_{i=1}^{2N}$,
the density of these eigenvalues without micro-alloying as $D^0(\lambda)$ and the density of states with a given concentration
of M particles as $D^M(\lambda)$ we show in Fig.~\ref{DOS} the measured density of states for models A and B.
To the untrained eye, the difference in the density of states seem slight. To make them more obvious, we show in
Fig.~\ref{diff} the {\em difference} between $D^M(\lambda)$ and $D^0(\lambda)$.
These plots provide us with valuable insights. Thus for example in Fig.~\ref{diff} upper panel we see that the density
of states at very small $\lambda$ is enhanced with the increase of the of the concentration of $M$ particles. In other
words, the addition of the M particles enhances the excess modes that are usually related to disorder and plasticity.
Since the non-affine term in the shear modulus is proportional to $\B H^{-1}$ it is relatively sensitive to the
increase in density of small eigenvalues, and hence the increase in the non-affine term. The Born term, as said above, is expected to increase when there is a significant amount of crystalline order. Crystalline order should
be detected by the increase in the density of states around the Debye cutoff (which is the value of $\lambda$ associated
with vibrations on the scale of the inter-particle distance). As explained above, in model A this is
not the case, as we see the very modest change in the density of states around the Debye cutoff $\lambda_D\equiv
8\pi\mu_B\approx 600$. Visual observation of the images in Figs.~\ref{struct} and \ref{voronoi} support this
concluson. On the other hand the fairly large crystalline patches formed in model B find a direct
correlation with the very large increase in the density of states around the Debye cutoff at larger values of $\lambda$.
This is the main reason for the significant increase in the Born term, leading to the eventual increase
in the shear modulus in model B.
\section{Summary and Conclusions}
\label{conc}
In summary, we pointed out that adding a small concentration of foreign particles to a given glass opens
up a rather huge parameter space that is not easy to systematize and organize. We demonstrated using two
simple examples in which the only difference was that the marked particles attracted preferentially the
larger particles or the smaller particles resulted in opposite tendencies in the mechanical properties.
We showed that the changes in the shear modulus in AQS conditions can be fully understood on the basis of
microscopic theory, for which the Hessian, its eigenvalues and the density of state of the latter provide
the correct parlance for understanding the observations. It is important however to understand that the
changes in the density of states are subtle, and difficult to predict a-priori. This is the fundamental reason
for the relatively haphazard and unpredictable consequences of micro-alloying. It is possible that by
studying carefully the glass forming properties of mixtures of constituents and finding their {\bf Eutectic
point} one can systematically attempt to improve mechanical properties by lowering this point \cite{13DJ}.
Studies of this line of thought will be presented elsewhere.

\acknowledgments
This work had been supported in part by the German Israeli Foundation, by an ERC 'ideas' grant `STANPAS'
and by the Israel Science Foundation.

\end{document}